# BIOCUBE: A MULTIMODAL DATASET FOR BIODIVERSITY RESEARCH

*Stylianos Stasinos**

Amazon
EU INTech
16 Rue Edward Steichen, 2540,
Luxembourg, Luxembourg

*Martino Mensio, Elena Lazovik, Athanasios Trantas*

Dutch Organization for Applied Scientific Research
TNO - ICT, Strategy & Policy
Anna van Buerenplein 1, 2595 DA, Den Haag,
The Netherlands

## ABSTRACT

Biodiversity research requires complete and detailed information to study ecosystem dynamics at different scales. Employing data-driven methods like Machine Learning is getting traction in ecology and more specific biodiversity, offering alternative modelling pathways. For these methods to deliver accurate results there is the need for large, curated and multimodal datasets that offer granular spatial and temporal resolutions. In this work, we introduce BioCube, a multimodal, fine-grained global dataset for ecology and biodiversity research. BioCube incorporates species observations through images, audio recordings and descriptions, environmental DNA, vegetation indices, agricultural, forest, land indicators, and high-resolution climate variables. All observations are geospatially aligned under the WGS84 geodetic system, spanning from 2000 to 2020. The dataset is available at https://huggingface.co/datasets/BioDT/BioCube, the acquisition and processing code base at https://github.com/BioDT/bfm-data.

## 1. INTRODUCTION

Biodiversity is undergoing rapid transformation due to human-induced environmental change, land-use shifts, and climate variability. Monitoring these changes at scale requires comprehensive datasets that not only capture singular modalities like species presence, but also contextual environmental information. However, most available biodiversity datasets are limited either to observational records or specific modalities such as imagery or genetic sequences, often lacking the necessary integration across environmental, spatial and temporal dimensions.

Recent advances in Digital Twins (DTs), Machine Learning (ML) and Earth Observation (EO) technologies have opened new avenues for ecological forecasting and biodiversity assessment. Yet, the full potential of these approaches is often hindered by challenges like fragmented data landscapes, inconsistent resolutions or modality gaps [1]. In response to these challenges we have engineered a multimodal dataset that provides a foundation for building scalable models that can be used for biodiversity monitoring, conservation planning and ecological forecasting at both global and local scales.

A series of ecology and biodiversity specialized datasets have recently emerged like BIOSCAN-5M [11] that contains over 5 million specimens of insects along with images, DNA barcode sequences, taxonomy, geographic information. Species distribution modeling is the focus of GeoLifeClef [4] dataset by merging 1.9 million plant and animal observations with high resolution remote sensing imagery, land cover and climate variables. In a similar direction, GeoPlant [7] provides over 5 million plant occurrence records across Europe, positivity enriched with Sentinel-2 satellite imagery and 20 years of climate time-series to support high-resolution spatial biodiversity observations. However, these datasets are far from containing enough diversified parameters to cover current needs in ecology. More specific, none of these datasets jointly integrates images, audio, eDNA, land, agriculture, conservation status, and climate variables. This gap motivated the construction of BioCube as a more diversified and holistic dataset.

The rest of the paper is organized as follows. Section 2 involves the methodology used, including data acquisition, preprocessing, and integration. In Section 3 the resulting data, its coverage, composition of modalities, and quality of the data are described. In Section 4, strengths, limitations, and open challenges regarding construction of large-scale biodiversity datasets that can be applied in ML are addressed. Finally, the prospect of the importance of BioCube to biodiversity research and ecological forecasting is discussed in Section 5.

## 2. METHOD

Latest ML methods like Foundation Models require large, well-curated, modality-rich datasets [2]. Accordingly, we assembled data from diverse sources, combining climate vari-

*Work performed during internship at TNO

ables, species observations, land indicators, and conservation records, as listed in Table 1. Acquisition used both API and file-based ingests to ensure scalability and spatiotemporal coverage.

### 2.1. Data Sources

BioCube dataset integrates data from several sources:

- **Climate Variables:** ERA5 hourly global reanalysis data, such as temperature, wind, pressure, and humidity, both in surface and atmospheric layers, obtained from the Copernicus Climate Data Store (CDS) [6].
- **Species Observations:** Images and metadata, such as taxonomy, geolocation, and timestamps collected from iNaturalist [1] and iNat2021 [5]. These datasets provide direct presence evidence.
- **Acoustic Data:** Bird vocalizations and metadata retrieved from Xeno-Canto [2] and from Xeno-Canto in GBIF [10], crucial for species monitoring where visual observation is difficult.
- **Species Descriptions and Conservation Status:** Textual records describing habitats, traits from Map of Life [9] and IUCN Red List [3], including red list index values and threat categories. The Red List is a global reference for extinction risk, with categories ranging from Extinct (EX) to Least Concern (LC).
- **Species Distribution:** Data is derived from the Living Planet Index [4], which aggregates population trends of species globally.
- **Environmental DNA (eDNA):** Genetic barcode sequences obtained from the Barcode of Life Data System (BOLD) [8].
- **Land and Vegetation Indicators:** NDVI sourced from Copernicus Land Services [5], and forest cover, land and agricultural indicators from The World Bank [6].

### 2.2. Acquisition Methods

API-based acquisition had a focus on the dynamic and real-time data retrieval. ERA5 climate variables were obtained through CDS using bounding boxes and temporal filters, with batch processing. Species data, including images, taxonomy, and geolocation, were collected via the iNaturalist API, while bird vocalizations were retrieved from the Xeno-Canto API, based on quality and location filters. Environmental DNA (eDNA) was sourced from the BOLD Systems API, and species descriptions together with the threat categories (e.g., IUCN Red List status) were accessed using the Map of Life API. To ensure efficiency and data integrity, we have implemented independent API modules to promote scalability and flexibility.

File-based acquisition provided an access to static and historical datasets, adding essential temporal depth and spatial coverage. The Living Planet Index (LPI) has contributed annual species distribution data from 1950 to 2020. NDVI products from Copernicus Land Services supplied vegetation indices recorded every 10 days at 1 km resolution, resampled to 0.25° grids for consistency. Land-use indicators, including arable land, irrigated areas, cropland extent, and forest cover, were sourced from the World Bank for the years 1961 to 2021. Offline datasets such as iNat2021 (2.7 million labeled images) and archived Xeno-Canto audio recordings accessed via GBIF further enriched the dataset. The complete file sizes and metadata can be found in Table 2.

### 2.3. Preprocessing

The obtained data could not be used in its raw format. Specific preprocessing steps detailed below needed to ensure consistency, quality and compatibility across modalities, while serving as a foundational component in the construction of structured data **batches** or **cubes** used for downstream modelling tasks. The preprocessing methods were performed during the dataset preparation phase and are crucial for generating uniform and high-quality inputs, harmonised to a 0.25° WGS84 geodesic coordinate grid, and temporally aligned to daily or monthly intervals.

- **Audio:** Silence removal, noise reduction (spectral gating), resampling, MFCC extraction.
- **Image:** Denoising, resizing, cropping.
- **Text:** Stopword and punctuation removal, stemming, lemmatisation, BERT embeddings and bag-of-words transformation.
- **eDNA:** Sequence filtering, k-mer vectorization and normalization.
- **Climate and Land Data:** Missing data interpolation, normalisation and temporal aggregation of variables such as temperature, wind, and pressure.

While end-to-end neural networks now dominate modern image, audio, and text analysis, we deliberately included traditional feature extraction methods (e.g., MFCC, TF-IDF) to ensure reproducibility and to provide baselines for researchers employing classical ML methods.

## 3. RESULTS

To construct unified dataset for biodiversity, we have acquired and curated multimodal data from multiple sources as a first step. Then the collected data available at Table 3 has been integrated into a structured species dataset with the following fields:

[1] https://www.inaturalist.org
[2] https://xeno-canto.org
[3] https://www.iucnredlist.org
[4] https://www.livingplanetindex.org/
[5] https://land.copernicus.eu/en/products/vegetation/normalised-difference-vegetation-index-v3-0-1km
[6] https://data.worldbank.org/indicator

**Table 1**. Overview of data modalities and variables included in the dataset.

| Modality | Source | Variables |
|---|---|---|
| **Surface Climate** | Copernicus (ERA5) | 2m temperature, 10m wind (u/v), mean sea-level pressure |
| **Atmospheric Variables** | Copernicus (ERA5) | Geopotential, temperature, humidity, wind (13 pressure levels: 50–1000 hPa) |
| **Single-Level Variables** | Copernicus (ERA5) | Land-sea mask, surface geopotential, soil type |
| **Species Observations** | iNaturalist, GBIF, Xeno-Canto | Images, audio, coordinates, timestamp, taxonomy |
| **Descriptions** | Map of Life | Text descriptions (behavior, habitat) |
| **eDNA** | BOLD Systems | DNA sequences (ATCG), taxonomic identifiers |
| **Distribution Trends** | Living Planet Index | Annual species occurrence and population trends (1950–2020) |
| **Red List Index (RLI)** | IUCN / Map of Life | Extinction risk index (0–1), categories: EX, EW, CR, EN, VU, NT, LC |
| **NDVI** | Copernicus Land (SPOT, PROBA-V) | Vegetation index values (-1 to 1), 10-day temporal resolution, 1 km spatial resolution |
| **Agri/Forest Indicators** | World Bank | Arable land, irrigated land, cropland area, forest cover, total land area |

**Table 2**. Data Sources by File Count and Total Size

| Data Source Name | Total Files | Total Size (GB) |
|---|---|---|
| Climate Variables | 24,510 | 160 |
| Species Observations | 51,918 | 52 |
| Acoustic Data | 43,511 | 104.4 |
| Species Descriptions | 20,593 | 0.005 |
| Environmental DNA | 16,257 | 0.1 |
| Species Destribution | 4,922 | 0.03 |
| Land Indicators | 7 | 0.0001 |
| Species Conversion Status | 1 | 0.011 |
| Vegetation Indicators | 258 | 88 |

- Species Identification: Species, Phylum, Class, Order, Family, Genus
- Location and Time: Latitude, Longitude, Timestamp
- Multimodal Inputs: Image, Audio, eDNA, Description, Redlist, Distribution

**Table 3**. Statistics of the Species Folder Contents

| Category | Count |
|---|---|
| Total number of Species | 40,282 |
| Species with eDNA, no images, no audios | 15,064 |
| Species with images, no audio, no eDNA | 16,630 |
| Species with images and audio, no eDNA | 1,849 |
| Species with audio, no images, no eDNA | 2,772 |
| Species with images and eDNA, no audio | 738 |
| Species with audio and eDNA, no images | 182 |
| Species with all modalities | 273 |

These records are extracted from over 40,000 species folders, each containing varying combinations of modalities. To efficiently extract relevant data, we implemented a folder filtering mechanism based on a hash-table-inspired approach. Each folder is being treated as a unique bucket, and its internal CSV files (image, audio, eDNA, etc.) are scanned for timestamps. Only folders containing at least one timestamp within the target date range (2000–2020) are selected for further processing. This has minimised memory usage and accelerated BioCube's construction time by avoiding unnecessary I/O on irrelevant folders.

**Image-Audio Matching:** When both images and audio were available, we matched them by averaging their metadata; latitude, longitude, timestamp, and paired them to maximise spatiotemporal alignment. Additionally, species-level data such as taxonomy and distribution are matched using the closest year and location for each sample.

**Efficient Storage:** All data is stored in Apache Parquet format to optimize I/O operations. Tensors are serialised (as base64-encoded arrays), and each sample is assigned a `unique_id` to avoid duplication. Latitude and longitude values are rounded to a 0.25-degree resolution to align with other datasets such as climate and land-use data. If different coordinates systems were found, we transformed them to WGS84 format. The data are saved incrementally after processing each folder, enabling scalable and resilient processing.

Land indicators (agriculture, forest cover, NDVI) required additional preprocessing. Because several sources report only country-level values, we extracted country bounding boxes and interpolated these to a spatial grid to align with species-level data. NDVI at 1 km monthly resolution was harmonised separately from annual forest and agriculture statistics obtained from the World Bank and Copernicus services. Table 4 summarises the total values only for Europe.

**Table 4**. Summary of Environmental Indicators (Europe)

| **Indicator Type** | **Total Values** | **Countries** |
|---|---|---|
| Agricultural (Arable) | 2,311,390 | 42 |
| Agricultural (Irrigated) | 411,482 | 33 |
| Cropland Area | 2,276,021 | 38 |
| Forest Cover | 1,285,834 | 44 |
| Land Area | 852,248 | 44 |
| NDVI (Vegetation Index)* | 15,929,016 | 44 |

* NDVI values are recorded **monthly**, while all other indicators are reported **annually**.

## 4. CONCLUSION AND DISCUSSION

BioCube marks a big step towards progressing biodiversity research through its complete multimodal analysis of fine-grained environmental and ecological data across the globe. BioCube connects species observations including imagery, audio recordings, environmental DNA data as well as descriptive information with precise climate, land-use data, vegetation measurements and conservation metrics to fill a significant research gap between singular modality datasets. Its main strengths lie in the breadth of data types, global geospatial alignment, and open availability. These enable research in species monitoring, conservation planning, and ecological forecasting. Still, some limitations remain, like NaN and missing values, taxonomic and geographic biases, reliance on legacy feature extraction, and as well as incomplete modal or spatial representation. Overall, BioCube is a scalable step toward developing biodiversity foundation models [2], supporting hybrid experiments now, while future work should expand modalities, improve coverage, and add real-time data streams.

## 5. ACKNOWLEDGMENTS

This study has received funding from the European Union's Horizon Europe research and innovation programme under grant agreement No 101057437 (BioDT project, https://doi.org/10.3030/101057437). Views and opinions expressed are those of the author(s) only and do not necessarily reflect those of the European Union or the European Commission. Neither the European Union nor the European Commission can be held responsible for them. This work used the Dutch national e-infrastructure with the support of the SURF Cooperative using grant no. EINF-10148.